\documentclass[aip,graphicx, amsmath,amssymb,reprint]{revtex4-1}
\usepackage{graphicx}
\usepackage{bm}
\usepackage{ulem}
\usepackage{color}

\def\jnl@style{\it}
\def\aaref@jnl#1{{\jnl@style#1}}

\def\aaref@jnl#1{{\jnl@style#1}}

\def\aj{\aaref@jnl{AJ}}                   
\def\araa{\aaref@jnl{ARA\&A}}             
\def\apj{\aaref@jnl{ApJ}}                 
\def\apjl{\aaref@jnl{ApJ}}                
\def\apjs{\aaref@jnl{ApJS}}               
\def\ao{\aaref@jnl{Appl.~Opt.}}           
\def\apss{\aaref@jnl{Ap\&SS}}             
\def\aap{\aaref@jnl{A\&A}}                
\def\aapr{\aaref@jnl{A\&A~Rev.}}          
\def\aaps{\aaref@jnl{A\&AS}}              
\def\azh{\aaref@jnl{AZh}}                 
\def\baas{\aaref@jnl{BAAS}}               
\def\jrasc{\aaref@jnl{JRASC}}             
\def\memras{\aaref@jnl{MmRAS}}            
\def\mnras{\aaref@jnl{MNRAS}}             
\def\pra{\aaref@jnl{Phys.~Rev.~A}}        
\def\prb{\aaref@jnl{Phys.~Rev.~B}}        
\def\prc{\aaref@jnl{Phys.~Rev.~C}}        
\def\prd{\aaref@jnl{Phys.~Rev.~D}}        
\def\pre{\aaref@jnl{Phys.~Rev.~E}}        
\def\prl{\aaref@jnl{Phys.~Rev.~Lett.}}    
\def\pasp{\aaref@jnl{PASP}}               
\def\pasj{\aaref@jnl{PASJ}}               
\def\qjras{\aaref@jnl{QJRAS}}             
\def\skytel{\aaref@jnl{S\&T}}             
\def\solphys{\aaref@jnl{Sol.~Phys.}}      
\def\sovast{\aaref@jnl{Soviet~Ast.}}      
\def\ssr{\aaref@jnl{Space~Sci.~Rev.}}     
\def\zap{\aaref@jnl{ZAp}}                 
\def\nat{\aaref@jnl{Nature}}              
\def\iaucirc{\aaref@jnl{IAU~Circ.}}       
\def\aplett{\aaref@jnl{Astrophys.~Lett.}} 
\def\apspr{\aaref@jnl{Astrophys.~Space~Phys.~Res.}}
\def\bain{\aaref@jnl{Bull.~Astron.~Inst.~Netherlands}} 
\def\fcp{\aaref@jnl{Fund.~Cosmic~Phys.}}  
\def\gca{\aaref@jnl{Geochim.~Cosmochim.~Acta}}   
\def\grl{\aaref@jnl{Geophys.~Res.~Lett.}} 
\def\jcp{\aaref@jnl{J.~Chem.~Phys.}}      
\def\jgr{\aaref@jnl{J.~Geophys.~Res.}}    
\def\jqsrt{\aaref@jnl{J.~Quant.~Spec.~Radiat.~Transf.}}
\def\memsai{\aaref@jnl{Mem.~Soc.~Astron.~Italiana}}
\def\nphysa{\aaref@jnl{Nucl.~Phys.~A}}   
\def\physrep{\aaref@jnl{Phys.~Rep.}}   
\def\physscr{\aaref@jnl{Phys.~Scr}}   
\def\planss{\aaref@jnl{Planetary~and~Space~Sciences}}   
\def\procspie{\aaref@jnl{Proc.~SPIE}}   

\begin{document}




\preprint{}

\title{Influence of the dissipation mechanism on collisionless magnetic reconnection in symmetric and asymmetric current layers} 








\author{Nicolas Aunai}
\email[]{nicolas.aunai@nasa.gov}
\affiliation{Space Weather Laboratory, Code 674, NASA Goddard Space Flight Center, Greenbelt, Maryland 20771, USA}

\author{Michael Hesse}
\affiliation{Space Weather Laboratory, Code 674, NASA Goddard Space Flight Center, Greenbelt, Maryland 20771, USA}

\author{Carrie Black}
\affiliation{Space Weather Laboratory, Code 674, NASA Goddard Space Flight Center, Greenbelt, Maryland 20771, USA}

\author{Rebekah Evans}
\affiliation{Space Weather Laboratory, Code 674, NASA Goddard Space Flight Center, Greenbelt, Maryland 20771, USA}

\author{Maria Kuznetsova}
\affiliation{Space Weather Laboratory, Code 674, NASA Goddard Space Flight Center, Greenbelt, Maryland 20771, USA}

\date{\today}

\begin{abstract}
 Numerical studies implementing different versions of the collisionless Ohm's law have shown a reconnection rate insensitive to the nature of the non-ideal mechanism occurring at the X line, as soon as the Hall effect is operating. Consequently, the dissipation mechanism occurring in the vicinity of the the reconnection site in collisionless systems is usually thought not to have a dynamical role beyond the violation of the frozen-in condition. The interpretation of recent studies have however led to the opposite conclusion that the electron scale dissipative processes play an important dynamical role in preventing an elongation of the electron layer from throttling the reconnection rate. This work re-visits this topic with a new approach. Instead of focusing on the extensively studied symmetric configuration, we aim to investigate whether the macroscopic properties of collisionless reconnection are affected by the dissipation physics in asymmetric configurations, for which the effect of the Hall physics is substantially modified. Because it includes all the physical scales a priori important for collisionless reconnection (Hall and ion kinetic physics) and also because it allows one to change the nature of the non-ideal electron scale physics, we use a (two dimensional) hybrid model. The effects of numerical, resistive and hyper-resistive dissipation are studied. In a first part we perform simulations of symmetric reconnection with different non-ideal electron physics. We show that the model captures the already known properties of collisionless reconnection. In a second part, we focus on an asymmetric configuration where the magnetic field strength and the density are both asymmetric. Our results show that contrary to symmetric reconnection, the asymmetric model evolution strongly depends on the nature of the mechanism which breaks the field line connectivity. The dissipation occurring at the X line plays an important role in preventing the electron current layer from elongating and forming plasmoids. 
\end{abstract}

\pacs{}

\maketitle 

\section{Introduction}
Magnetic reconnection is a universal plasma phenomenon releasing magnetic energy into thermal and bulk kinetic energy\citep{Priest:2000wm,Birn:2007vz,Yamada:2010im}. Since reconnection is thought to be at the root of very fast energy release events, a major issue is the understanding of the physical mechanisms controling the rate at which the magnetic flux is being reconnected, and at which scale they operate. Reconnecting field lines requires that all the plasma species violate the frozen-in condition. In collisionless systems, this leads the structures to thin down to scales where finite Larmor radius effects become important. The electrons are the lightest species, they will therefore control the scale at which reconnection ultimately occurs in such systems, but not necessarily the scale that controls the overall rate. The puzzle then consists in understanding whether the electron scale physics, the (larger) ion scale physics or a coupling between the two, controls the overall dynamics of the reconnection process and how.

Modeling magnetic reconnection with different simplified versions of the collisionless Ohm's law has led to great progress in understanding which mechanism plays a pivotal role in the collisionless reconnection process. It is now generally accepted that the Hall physics plays a crucial role in enabling a fast regime, \textit{i.e.} a rate that is large enough for reconnection to be consistent with the phenomena it is considered to be a key ingredient of. Multiple numerical studies have indeed shown that including the Hall effect in reconnection models results in a large rate, which is moreover quite insensitive to the non-ideal electron dynamics occurring close to the reconnection site\citep{2001JGR...106.3715B,2007PhRvL..99o5002S}. 

A paradigm, generally called the \textit{Hall reconnection model}, has been proposed to explain why the Hall effect alone results in these two facts\citep{2001JGR...106.3715B,1998GeoRL..25.3759S,2001JGR...106.3759S,1994GeoRL..21...73M,2001PhRvL..87s5004R,2008PhPl...15d2306D}. First, the Hall effect, owing to its dispersive properties, is thought to make the outward electron flux insensitive to the mechanism breaking the field line connectivity. A Sweet-Parker\citep{1957JGR....62..509P}-like reasoning then leads to the conclusion that the reconnection rate must remain unaffected by any parameter altering the physics at this scale, no matter what it actually is as long as it falls well below the ion inertial length. The second consequence of the Hall effect is the propagation of Hall fields along the separatrices, which opens the exhaust by heating the ions and accelerating them to maintain a large downstream flux, independent of the system size.

Understanding whether collisionless reconnection is sensitive or not to the electron scale physics occurring at the X line or is entirely controlled by the Hall effect alone, however remains an open question.  Recent simulations\citep{2006PhPl...13g2101D,2007GeoRL..3413104K} of domains larger than before and/or with open boundaries, have revealed an elongation of the electron current layer in the downstream direction. Evidence was found to support the idea of a relationship between this elongation and the concomitant decrease of the reconnection rate. Two mechanisms were proposed to explain the cessation of elongation. One mechanism is the triggering of a secondary tearing instability, which, by breaking the current sheet into multiple plasmoids, would dynamically prevent further elongation of the layer, and at the same time, make reconnection a highly unsteady process. A second mechanism would be the balance in the downstream direction of the outward propagation of magnetic flux with the dissipation occurring at the X line, which could occur for long periods of time and make the reconnection process appearing as quasi-steady. The question of what would cause an unbalance between these two processes and then result in the elongation of the current sheet has however not been addressed. 

On the other hand, other fully kinetic simulations\citep{2008PhPl...15d2306D,2007PhRvL..99o5002S}, in quite similar initial configurations, have shown a fast and steady reconnection rate and insensitive to variations of the electron mass, supporting the Hall reconnection paradigm. As a result of these simulations, it was concluded that the extent of the electron layer in the downstream direction is controlled by the Hall effect alone and its ability to rotate the reconnected magnetic field in the out-of-plane direction with a speed depending on the scale. Furthermore, Hall MHD simulations without explicit dissipation physics, i.e. with only numerical dissipation, have revealed a fast reconnection rate, similar to the one measured in fully kinetic systems, and moreover independent on the mesh resolution\citep{2001JGR...106.3737B,1999PhPl....6.1781H}. These last results are consistent with the electron dissipative physics playing no role and give the Hall effect the full control of the reconnection dynamics.

Finally, recent theoretical investigations\citep{2011SSRv..tmp...10H} have shed new lights on the physical role of the electron dissipation scale mechanisms. The structure of the electron full pressure tensor at the X line has been explained from the kinetic viewpoint to be related either to the electron bounce motion around reversed field lines, or to the mixing of cold/slow electrons entering the layer with hot/fast electrons leaving it, depending on the coplanarity of the macroscopic magnetic configuration. The theory showed that this non-gyrotropic contribution to the electron momentum can be related to a viscous effect from the fluid viewpoint, the off-diagonal components of the pressure tensor being linked to the second derivative of the electron bulk velocity. Physically, the role of the reconnection electric field can thus be understood as the way to sustain the current density required at the X line to maintain the magnetic shear, which would otherwise be dissipated by the diffusion of electrons accelerated in the layer and leaving it.

In this paper, we aim to study further the sensitivity of collisionless reconnection to the dissipation physics. Since the classic reconnection setup seems to lead to rather different conclusions, we adopt a different approach, and decide to investigate how sensitive collisionless reconnection is to the dissipation physics in a more general, asymmetric, configuration too, i.e. where the reconnecting current sheet is no longer separating two identical plasmas, in terms of density and magnetic field strength. The initial perfect symmetry of the current sheet, extensively used to initialize numerical models, is indeed hardly expected anywhere, but, to some extent, at the Earth magnetotail. Besides, the few studies which have been so far focused on asymmetric reconnection have shown that the structure of the Hall fields, on which relies the Hall paradigm, is significantly modified\citep{Pritchett:2008ef,2009JGRA..11411210P,2009PhPl...16h0702P,2008AnGeo..26.2471T,2003JGRA..108.1218S,2000JGR...10523179N,1997GeoRL..24.3145L,2011SSRv..158..119M,2010JGRA..11510223M}. It is therefore important to understand how these systems respond with regard to the current debate on the role of the dissipative physics. We will focus on the role, in reconnection, of the dissipation, as defined from the fluid viewpoint, as any mechanism that is not considered as ideal and which enables field lines to change their connectivity. Previous understanding may lead to think that any dissipation effect coming after the Hall term in the Ohm's law, should leave the overall evolution of the reconnection process unchanged. This question does not need a fully kinetic code to be addressed, as the hybrid kinetic formalism, includes all the ingredients previously mentioned as part of the steady Hall reconnection paradigm, neglects electron non-ideal effects to which the paradigm attributes no dynamical role and let us control the dissipative physics by choosing different mechanisms and investigate their respective role. Any deviation from the standard steadiness and insensitivity to the dissipation mechanism would emphasize the need for a better understanding of reconnection and would allow one to identify key ingredients for its modeling. Furthermore, it is important to understand to what extent the use of a dissipation mechanism, versus another one, in a hybrid code, changes the macroscopic evolution of asymmetric reconnection, as these codes are the only one able to simulate large scale phenomena and include the appropriate ion dynamics at the same time. This paper does however not investigate whether the chosen dissipation models can be justified by specific underlying kinetic processes, and only looks, from the macroscopic viewpoint, whether any model results in the same overall evolution of reconnection or not, in symmetric and asymmetric configurations. The comparison of that overall evolution with a fully kinetic model is a topic explored in a separate paper\citep{aunai2013a}. The investigation of how the kinetic processes, occurring at the X line in asymmetric reconnection, can result, from the fluid viewpoint, in dissipation, and how it compares to the present dissipation mechanisms is also considered off-topic and should be investigated in the future using, this time, fully kinetic simulations.

In the second part of this paper, we will present in detail the numerical model used to perform this study. Then, the third part will be focused on hybrid simulations of symmetric reconnection with different dissipation mechanisms. This part will show to what extent our model recovers the already known properties of Hall reconnection. In a fourth part, we will investigate the role of dissipation in asymmetric reconnection. Finally the fifth part will summarize and discuss our findings.

\section{Numerical model}
The hybrid model we use solves the ion (assumed to be protons) kinetic dynamics using the particle in cell method\citep{Birdsall:2004vm}. Their motion is obtained from equation (\ref{eq:lorentz}), where $m_i$ is the proton mass, $\mathbf{E}$ and $\mathbf{B}$ are the electric and magnetic field, respectively.  This equation is solved using the Boris algorithm\citep{Birdsall:2004vm}. The electron population is assumed to be a fluid, meaning that the infinite chain of moments of their distribution function is truncated at the pressure $P_e$ order, for which an isothermal law is assumed (\ref{eq:isothermal}) where $T_e$ is the constant and uniform electron temperature. The electron bulk inertia is also neglected. At all locations and times, the electron fluid is assumed to have the same density as the proton density $n$, so that quasineutrality holds. The proton density itself is obtained from the proton velocity distribution (\ref{eq:density}). The electron momentum equation is used to calculate the electric field in the form of an Ohm's law (\ref{eq:ohm}), where $\mathbf{R}$ is a dissipation term. Consistently, the displacement current is neglected in Maxwell-Ampere's equation (\ref{eq:ampere}), where the definition of the current density $\mathbf{j} = en\left(\mathbf{v}_i - \mathbf{v}_e\right)$ has been used, with $e$ the elementary charge, $\mathbf{v}_i$ and $\mathbf{v}_e$ are the proton and electron bulk flows, respectively. Similarly to their density, the proton bulk flow is calculated from the velocity distribution (\ref{eq:meanvel}).  The magnetic field is evolved using Faraday's law (\ref{eq:faraday}).

\begin{eqnarray}
m_i\frac{d\mathbf{v}_{pi}}{dt} & = & e\left(\mathbf{v}_{pi}\times \mathbf{B} + \mathbf{E}\right)\label{eq:lorentz}\\
\mathbf{\nabla}\times\mathbf{B} & = & \mu_0en\left(\mathbf{v}_i - \mathbf{v}_e\right)\label{eq:ampere}\\
\mathbf{E} & = & -\mathbf{v}_i\times\mathbf{B} + \frac{1}{ne}\left(\mathbf{j}\times\mathbf{B}-\boldsymbol{\nabla} P_e\right) + \mathbf{R}\label{eq:ohm}\\
\frac{\partial \mathbf{B}}{\partial t} & = & -\mathbf{\nabla}\times\mathbf{E}\label{eq:faraday}\\
P_e & = & nk_BT_e\label{eq:isothermal}\\
n & = & \int f\left(\mathbf{r},\mathbf{v}_{pi}\right) d\mathbf{v}_{pi}\label{eq:density}\\
\mathbf{v}_i & = &\frac{1}{n} \int \mathbf{v}_{pi}f\left(\mathbf{r},\mathbf{v}_{pi}\right) d\mathbf{v}_{pi}\label{eq:meanvel}
\end{eqnarray}

The dissipation (term $\mathbf{R}$ in eq. (\ref{eq:ohm})) is alternatively chosen to be negligible ($\mathbf{R} = \eta\mathbf{j} \approx 0$) in a way that it is not able to prevent the collapse of the current sheet thickness down to the grid scale but sufficient to prevent the simulation from crashing due to accumulation of noise, resistive ($\mathbf{R} = \eta\mathbf{j}$) and non-negligible so that the current sheet thickness is controlled by Joule diffusion, or hyper-resistive ($\mathbf{R} = -\nu\boldsymbol{\nabla}^2\mathbf{j}$). The resistivity $\eta$ and the hyper-resistivity $\nu$ are both considered constant in time and spatially uniform. Classical resistivity is related to electron-ion collisions, whereas hyper-resistivity can be seen as a model of electron viscosity\citep{Biskamp:2005wz}. Other models for non-ideal terms exist but we have not implemented them for the following reasons. Localized resistivity is known to be the source of fast reconnection without the Hall term\citep{2001PhPl....8.4729B} and thus seems inadequate for this study. Moreover, unless a somewhat arbitrary and complicated model is chosen, it has a constant and not self-consistent scale and is spatially fixed, which can be problematic considering the possible drifts of the X line seen in asymmetric systems\citep{2003JGRA..108.1218S,2009JGRA..11411210P,Pritchett:2008ef}. Electron bulk inertia can also be non-negligible around the reconnection site. Usually, this term is implemented for numerical reasons in two-species models because it conveniently changes the whistler dispersion relation and prevents whistler waves to severely limit the CFL condition. However its standard implementation requires one to assume a uniform electron inertial length where it should be local\citep{2001JGR...106.3759S,2008GeoRL..3519102C}. Accordingly, it is unclear whether the electron mass appearing in such implementation has the same role as in real systems where it moreover also plays a role in the full pressure tensor. Let us note, on the other hand, that hyper-resistivity, being a viscous operator, shares remarkable similarities with the current understanding of the dissipative nature of symmetric electron current layer\citep{2011SSRv..tmp...10H}. It therefore appears as a reasonable and simple phenomenological model, although a more general, asymmetric, theory is needed.

Numerically, the equations for the electromagnetic fields are discretized with second order finite differences. The magnetic field components being defined at the same location, shifted by half a cell from all electric field components. Their time evolution is calculated with a predictor-corrector scheme. The mesh size is constant and uniform, and chosen so that the smallest physical scale in (\ref{eq:ohm}) is resolved. The time step is chosen so that the proton cyclotron period and the whistler propagation speed are resolved. Considering the thermal and bulk speeds in the presented simulations, these criteria guarantee, in particular, that no particle crosses an entire cell within a single time step. The results are presented in the plane $(x,y)$ and in dimensionless units. The particle density and the magnetic field are normalized by arbitrary quantities $n_0$ and $B_0$ respectively. Distances are normalized by the proton inertial length $\delta_i = V_A/\Omega_{ci}$, where $V_A$ and $\Omega_{ci}$ are the Alfven speed and the cyclotron frequency based on the density $n_0$ and magnetic field $B_0$.  The reconnection of magnetic field is initialized with a localized magnetic perturbation. For all the runs presented in this paper, the boundaries are periodic in the $x$ direction, and closed and perfectly conducting in the $y$ direction.

\section{Symmetric reconnection}
In this section, our goal is to show how symmetric reconnection is affected by the nature of the dissipation mechanism and to what extent these findings agree with previous ones before studying the more complicated case of asymmetric reconnection. To achieve this goal, we initialize our code with a symmetric tangential current layer. The magnetic field has initially only one component $B_x\left(y\right) = \tanh\left(\left(y-y_0\right)/\lambda\right)$, where $y_0$ is the middle of the domain in the $y$ direction, and $\lambda=0.5$ is the half thickness of the magnetic field reversal. The normalized density is chosen to be uniform and its value is set to unity. The asymptotic ratio of the thermal pressure and magnetic pressure $\beta$ is $1$. The proton temperature is chosen to be isotropic everywhere and is obtained from the pressure balance condition $nT + B^2/2\mu_0 = cst$. The electron temperature is set to $T_e=0.25$, is constant in time and spatially uniform. The protons are initially loaded as locally Maxwellian velocity distribution functions. This initial condition is not a Vlasov-Maxwell equilibrium and small magnetosonic waves are emitted from the current sheet within the first proton cyclotron period. These waves might modulate a bit the reconnection rate as they are reflected back to the current sheet from the closed top boundaries, but do not contribute to the presented results. Unless mentioned otherwise, the domain size in the $x$ direction is $x_m=250$ and $y_m=50$ in the $y$ direction, so that recirculation of plasma and waves do not affect the results within the time of interest. \\

We will study the results of four simulations differing only by the dissipative physics. The first two runs, labelled $S_{nd}^{a}$ and $S_{nd}^b$, aim to be  cases where the Joule dissipation is not strong enough to prevent the current sheet thickness to collapse down to the grid scale. The number of cells of run $S_{nd}^a$ in each direction is set to $\left(n_x,n_y\right) = \left(2500, 500\right)$. The spatial resolution is then $\left(\Delta x, \Delta y\right) = \left(0.1, 0.1\right)$. The run $S_{nd}^b$ has the same number of cells but the size of the domain in both directions is doubled, so that the mesh resolution is $0.2$ in both directions. This will allow us to see whether the resolution has an impact on the result in these dissipationless cases. We have used $\eta = 2\ 10^{-3}$. As a result of such a small resistivity combined with these mesh sizes, the current sheet collapses at the grid scale, meaning that doubling $\Delta y$ results in dividing the current density by a factor of 2 (the peak current density being given by $j_z \sim 1/\Delta y$). In this discrete simulated system, the breaking of field line connectivity is ensured by finite mesh effects and the thickness of the current sheet is obviously limited by the mesh size.

 In the third run $S_\eta$, all parameters are the same but the resistivity has been increased up to $\eta = 1.4\ 10^{-2}$. In the fourth run, we neglect the classical resistivity and choose the dissipative electric field to be $\mathbf{R} = -\nu\boldsymbol{\nabla}^2 \mathbf{j}$, with $\nu=3\ 10^{-4}$. In this last run, the resolution is doubled by increasing the number of cells by a factor 2 in the $y$ direction. The time step is set to $\Delta t = 0.001$ for all runs except run $S_{nd}^b$ for which it is $\Delta t = 0.01$. Table \ref{table:symmetricruns} summarizes the run parameters.\\

 \begin{table}

 \caption{\label{table:symmetricruns}Summary of the parameters}

 \begin{tabular}{|c|c|c|c|c|}
 \hline 
 \textbf{Name} & $\eta$ & $ \nu$  & $\Delta y$   \\
 \hline
$S_{nd}^a$ & $2\ 10^{-3}$ &   $0$ &  $ 0.1$\\
\hline 
$S_{nd}^b$ & $2\ 10^{-3}$ &   $0$ &  $ 0.2$\\
\hline 
$S_{\eta}$ & $1.4\ 10^{-2}$ & $0$ & $ 0.1$\\
\hline 
$S_{\nu}$ & $2\ 10^{-3}$  &  $3\ 10^{-4}$  &  $ 0.05$\\
\hline
 \end{tabular}

 \end{table}

\begin{figure}[htbp]
\begin{center}
\includegraphics[width=0.8\linewidth]{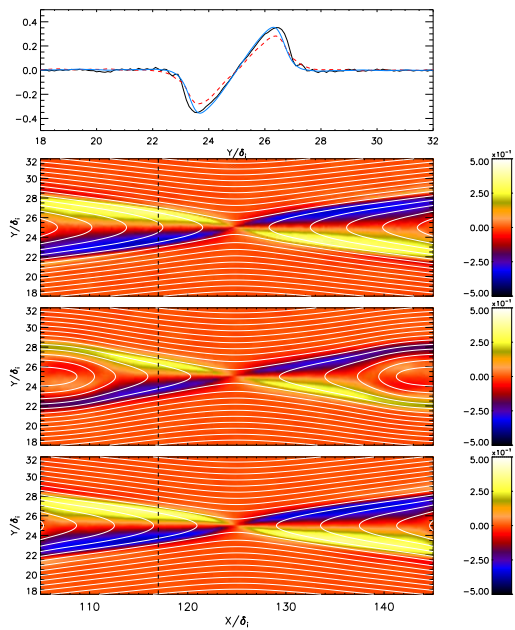}
\caption{\textbf{Top panel : } Out-of-plane component of the magnetic field measured as a function of $y$ at $x=117$ in $S_{nd}^a$ (black curve), $S_{\eta}$ (red curve) and $S_{\nu}$ (blue curve). \textbf{Then, from top to bottom : } Out-of-plane component of the magnetic field ($B_z$) for run $S_{nd}$ (first panel) at $t=50$ , run $S_{\eta}$ (second panel) at $t=57.5$ and run $S_{\nu}$ (last panel) at $t=52.5$. The snapshot are averaged over $\Omega_{ci}^{-1}$ around the indicated time, which is when all runs have reconnected the same amount of upstream magnetic flux. On each color plot, the black vertical dashed line shows the position of the cut represented on the top panel.}
\label{fig:bzsym}
\end{center}
\end{figure}

\begin{figure}[htbp]
\begin{center}
\includegraphics[width=\linewidth]{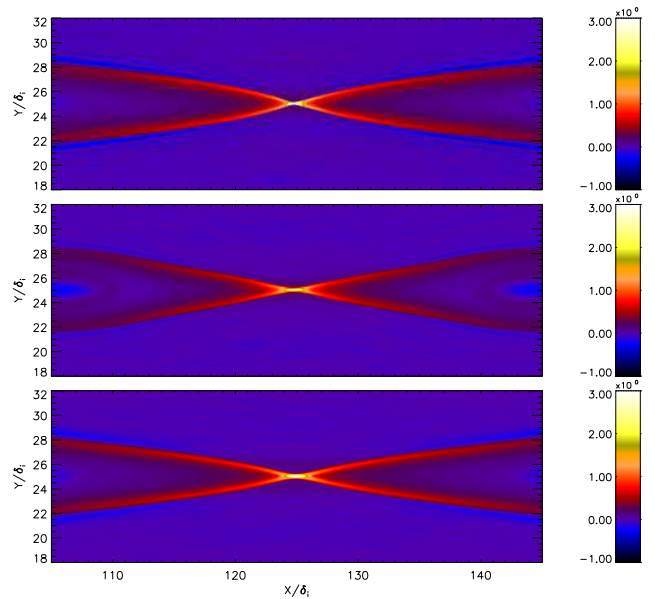}
\caption{Out-of-plane component of the electron bulk flow ($v_{ez}$) for run $S_{nd}$ (top panel), run $S_{\eta}$ (middle panel) and run $S_{\nu}$ (bottom panel). The snapshot are averaged over $\Omega_{ci}^{-1}$ around the same time as in fig \ref{fig:bzsym}.}
\label{fig:vezsym}
\end{center}
\end{figure}

\begin{figure}[htbp]
\begin{center}
\includegraphics[width=\linewidth]{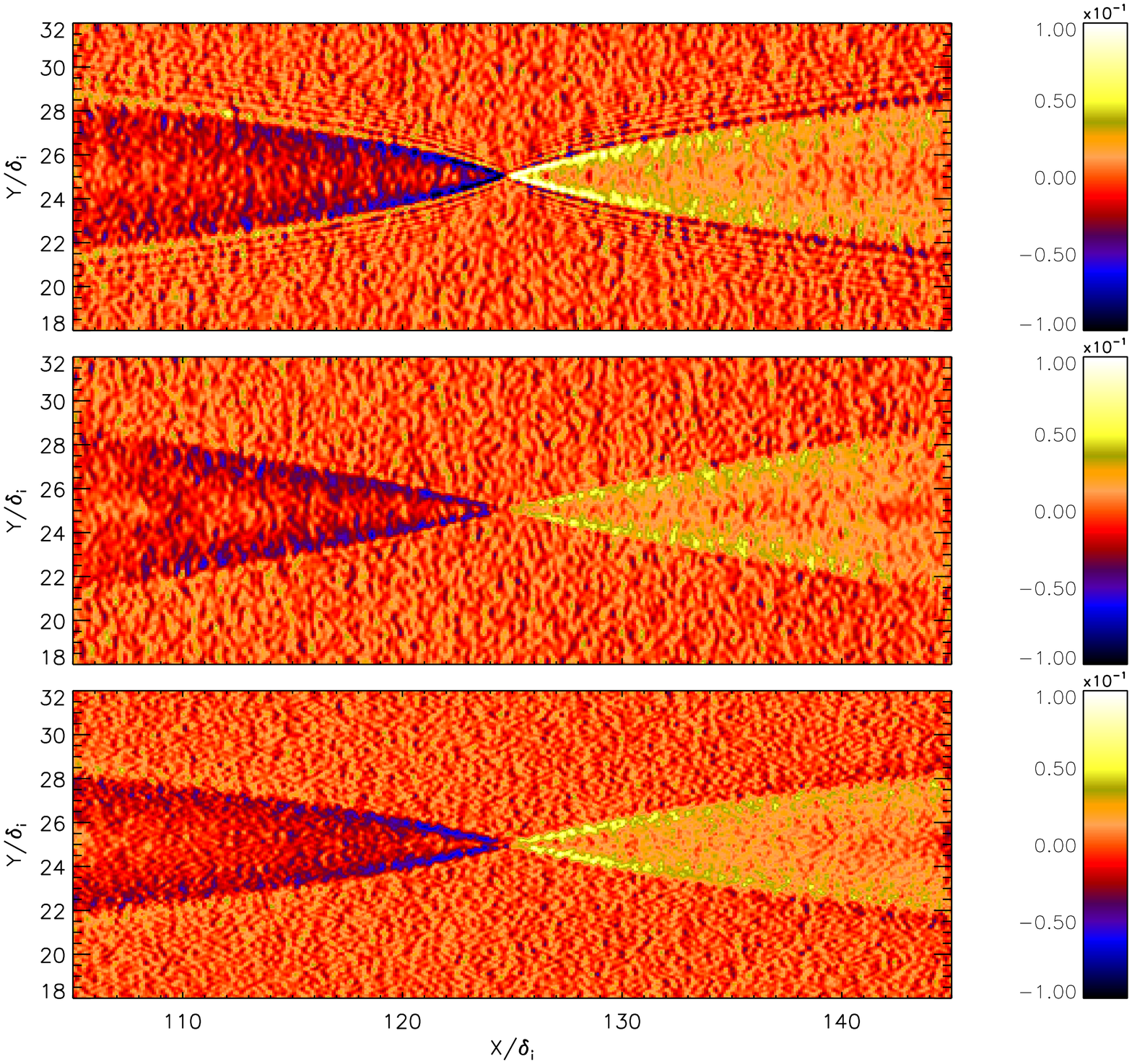}
\caption{$E_x$ electric field for run $S_{nd}$ (top panel), run $S_{\eta}$ (middle panel) and run $S_{\nu}$ (bottom panel). The snapshot are averaged over $\Omega_{ci}^{-1}$ around the same time as in fig \ref{fig:bzsym}.}
\label{fig:exsym}
\end{center}
\end{figure}

\begin{figure}[htbp]
\begin{center}
\includegraphics[width=\linewidth]{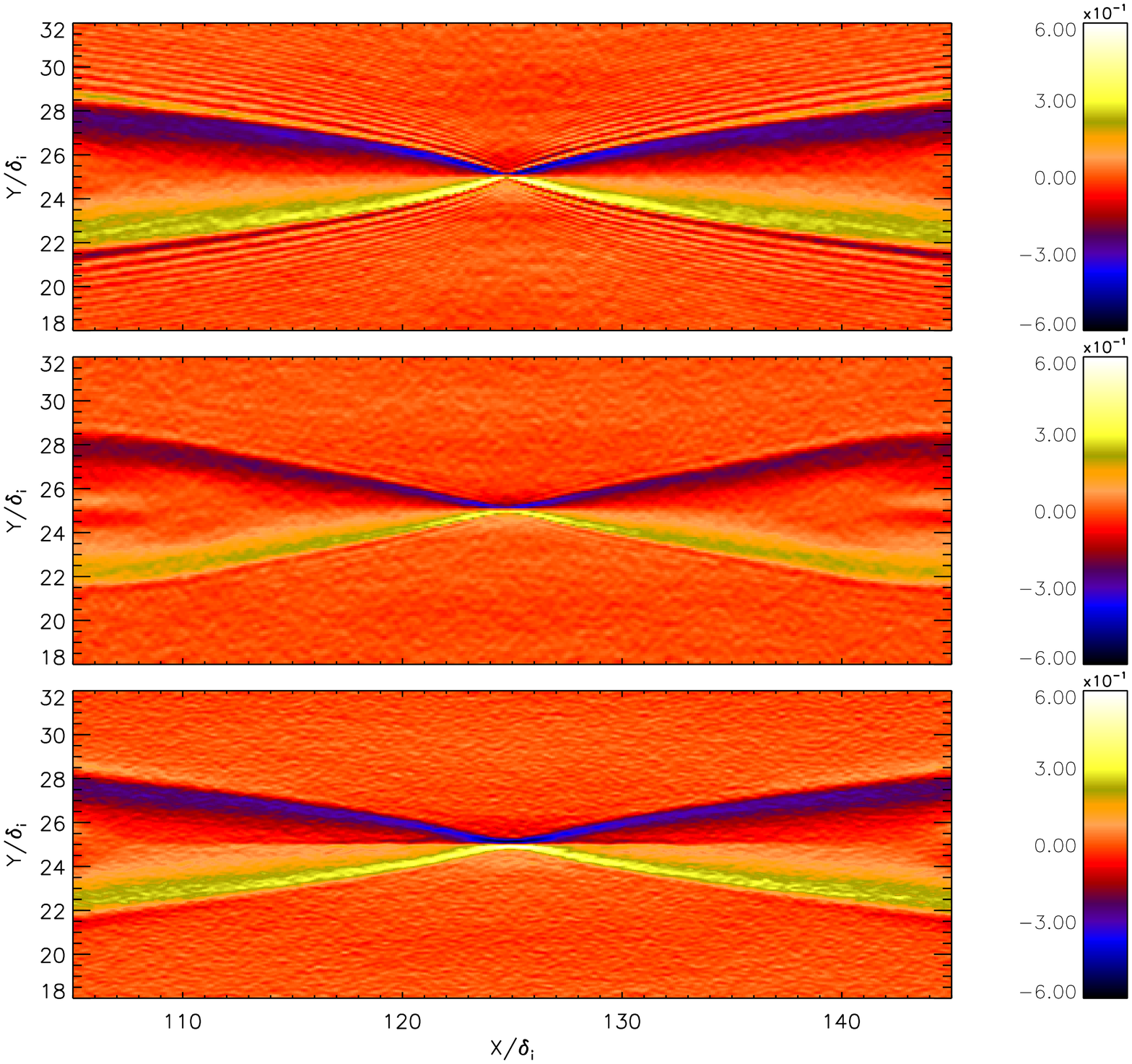}
\caption{$E_y$ electric field for run $S_{nd}^a$ (top panel), run $S_{\eta}$ (middle panel) and run $S_{\nu}$ (bottom panel). The snapshot are averaged over $\Omega_{ci}^{-1}$ around the same time as in fig \ref{fig:bzsym}.}
\label{fig:eysym}
\end{center}
\end{figure}

Figures \ref{fig:bzsym} to \ref{fig:eysym} show snapshots from the three symmetric simulations, $S_{nd}^a$, $S_{\eta}$ and $S_{\nu}$, at the time they have reconnected the same amount of magnetic flux. Run $S_{nd}^b$ is not shown to save space, but is found very similar to $S_{nd}^a$ and will be discussed later. On all these figures, the top panel represents the run $S_{nd}^a$ ($t=50$), the middle panel the run $S_\eta$ ($t=57.5$) and the run $S_\nu$ ($t=52.5$) is shown on the bottom panel. At first glance the three runs look very similar, suggesting that the dissipative term in eq. (\ref{eq:ohm}) does not play a critical role in the evolution of the system. A closer inspection reveals interesting details. Fig. \ref{fig:bzsym} shows the out-of-plane magnetic field component. As expected for symmetric collisionless reconnection, this component has a quadrupolar structure around the X point which does not extend upstream of the separatrices. This is due to the Hall rotation of the upstream in-plane field lines to the out-of-plane direction as they cross the separatrices. Looking at cuts in the $y$ direction on the top panel of Fig. \ref{fig:bzsym}, one can see that if the runs $S_{nd}^a$ and $S_\nu$ have very similar structure and values ($\approx 0.4$ as in previous works including fully kinetic PIC simulations\citep{2008PhPl...15d2306D,2001JGR...106.3827A,1998JGR...103.4547L,1998JGR...103.9165S,2001JGR...106.3759S,2009JGRA..11405111D}), the values of $B_z$ for the run $S_\eta$ are somewhat smaller ($\approx 0.25$).

Figure \ref{fig:vezsym} represents the out-of-plane electron bulk velocity $v_{ez}$. In the three simulations, $v_{ez}$ is strongly enhanced at the X point and, to a smaller extent, in the separatrix regions. These characteristics are also consistent with previous studies \citep{2009GeoRL..3607107M,2008PhPl...15d2306D, 2009JGRA..11405111D, 1998JGR...103.9165S}. In both runs $S_{nd}$ and $S_{\eta}$, the current sheet thickness is comparable to the grid spacing in the $y$ direction, whereas it is thicker in run $S_{\nu}$. This indicates that, for this resolution at least, the resistivity is not large enough to efficiently dissipate the incoming magnetic flux upstream of the X point. The Joule diffusion seems however to play a role in controlling the length of the layer, as it can be noticed that in run $S_{nd}^a$ it is close to the $x$ grid spacing whereas it appears longer in run $S_\eta$. This last point is more evident when observing the in-plane electric field.

Figures \ref{fig:exsym} and \ref{fig:eysym} show the $E_x$ and $E_y$ components of the electric field, respectively. They are zero in the regions upstream of the separatrices and are strongly peaked at their location. Peak values are $\approx 0.1$ and $\approx 0.5$ for $E_x$ and $E_y$, respectively, which is consistent with what is observed in other studies \citep{2008PhPl...15d2306D,2009JGRA..11405111D}. Both $E_x$ and $E_y$ are weaker in the run $S_{\eta}$ than in the two other runs. In run $S_{nd}^a$, $E_x$ goes up to the X point since the dissipation region has a length comparable to the grid spacing, in the other runs it starts a bit further and is yet another evidence that dissipation increases the length of the current layer. This feature can also be observed in fully kinetic simulations\citep{2008PhPl...15d2306D}. In contrast, $E_y$ is not zero in the current sheet except at symmetry line itself, and has a strong bipolar variation in the $y$ direction. This structure appears in run $S_\nu$, to a smaller extent in run $S_{\eta}$, but not in run $S_{nd}^a$, again because of the collapse of the electron current sheet down to the grid scale in both directions. The small waves seen in Fig. \ref{fig:vezsym} for run $S_{nd}^a$ can also be observed in both $E_x$ and $E_y$ components.

Although the Joule diffusion appears to be not strong enough to efficiently dissipate the incoming magnetic flux inside the electron current layer, the somewhat smaller values of the electromagnetic fields and electron flows observed in run $S_\eta$ indicate that it dissipates the energy at larger scales. The large scale diffusion is not seen in the dissipationless run $S_{nd}^a$, and not seen either in run $S_\nu$. This behavior is consistent with previous understanding of the effect of the Joule diffusion in collisionless magnetic reconnection\citep{2001JGR...106.3715B}.  In run $S_{nd}^a$, small waves can be observed upstream of the separatrices. Their absence in the two other runs and in fully PIC simulations suggests they are a spurious consequence of the lack of dissipation in the model.

\begin{figure}[htbp]
\begin{center}
\includegraphics[width=\linewidth]{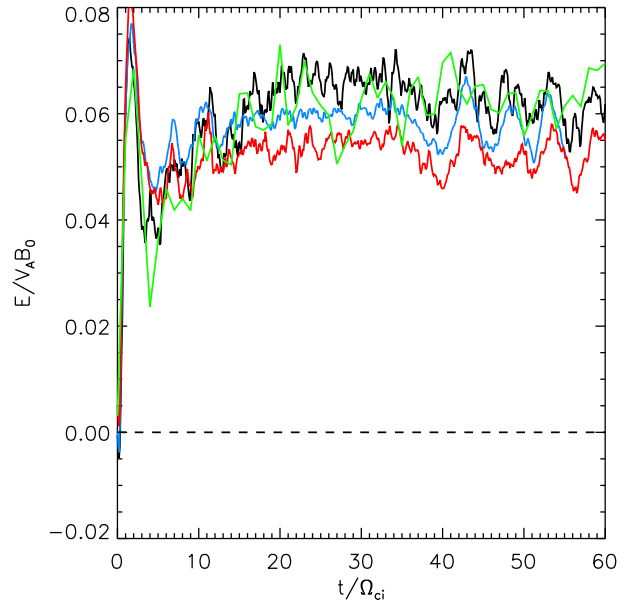}
\caption{Time evolution of the reconnection rates for the runs $S_{nd}^a$ (black solid), $S_{nd}^b$ (green), $S_{\eta}$ (dashed red) and $S_{\nu}$ (solid blue).}
\label{fig:symrates}
\end{center}
\end{figure}

Figure \ref{fig:symrates} shows the reconnection rates measured in all simulations. The rate is measured as the out-of-plane electric field $E_z$, averaged over a unit squared area centered on the magnetic flux saddle point, detected at each time. The black curve represents the reconnection rate of run $S_{nd}^a$. Its steady value is $E_{nd}\approx 0.065$, and goes to $\approx 0.13$ if normalized to the $V_{Aup}B_{up}$, the characteristic electric field at the edge of the averaging area. The green curve is the reconnection rate measured in run $S_{nd}^b$. Appart from the somewhat larger fluctuations and the different sampling rate, it is very similar and therefore suggests that the mesh resolution does not impact the reconnection process dramatically. The red dashed curve represents the reconnection rate measured in run $S_{\nu}$. Its value is $E_\nu \approx 0.06$, slightly below $E_{nd}$. The rate measured in run $S_{\eta}$ (blue line) is however significantly below, with a value $E_{\eta} \approx 0.055$. Note that the oscillatory period $t \in [45,55]$  corresponds approximately to the time for the waves emitted by the current sheet as a consequence of the non-equilibrium kinetic initial state, to return back at the center of the domain after having been reflected at the top boundaries. \\

\section{Asymmetric reconnection}
In this section we change our initial condition to include an asymmetry between both sides of the initial tangential current layer. We choose an initial condition that has already been used in previous works \citep{Pritchett:2008ef,2011PhRvL.106s5003Z} for the sake of comparison. It consists in a one dimensional tangential asymmetric current sheet, where both the magnetic field and the particle density balance the total pressure, leaving the electron and ion temperature initially uniform throughout the system. The density is given by (\ref{eq:densityinit}) and the magnetic field by (\ref{eq:magneticinit}), using the shape factor (\ref{eq:shapefactor}), centered at $y_0$, the middle of the domain in the $y$ direction, with a half-width $\lambda = 0.5$. A guide field of amplitude $B_{gf}=1$ has been added for more generality. Although we do not show them here to be more concise, we have performed other test simulations within a coplanar initial condition and have observed a very similar behavior. The initial plasma temperature is $T=3/2$ and the ion to electron temperature ratio is $T_i/T_e=5$. Particles are loaded as locally Maxwellian distribution functions. This initial condition is again not a Vlasov equilibrium and the current sheet initially evolves slightly toward a state closer to a self-consistent equilibrium while emitting some ion scale waves in the system. These waves have not been found to affect the results discussed here in any way, although a kinetic steady state would certainly be preferable for both physical and practical reasons\citep{2012PhPl...19b2108B}. This initial phase of asymmetric reconnection modeling will be addressed in further detail in a forthcoming study.

\begin{eqnarray}
n(y) & = & 1 - \frac{1}{3}\left(S\left(y\right) +S\left(y\right)^2\right)\label{eq:densityinit}\\
B_x(y) &  = & \frac{1}{2} + S\left(y\right)\label{eq:magneticinit}\\
S\left(y\right) & = & \tanh\left(\frac{y-y_0}{\lambda}\right)\label{eq:shapefactor}
\end{eqnarray}

As in the previous section, we performed simulations with grid scale reconnection, resistive dissipation and hyper-resistive dissipation and investigate whether this changes lead to substantial differences in the dynamical behavior of reconnection. Runs with only numerical dissipation will be labelled $A_{nd}^{a,b}$, runs with classical resistive dissipation will be labelled $A_\eta^{a,b}$, and those with hyper-resistive dissipation will be labelled $A_\nu^{a,b}$. For each case, we perform two simulations with two mesh resolutions. The following analysis is performed at times occurring before  $t=\tau_A$, where $\tau_A = x_m/V_A$ is the shortest Alfven travel time across the system. This guarantees that the periodicity of the system does not alter significantly our results. Table \ref{table:asymmetricruns} summarizes the run parameters.

 \begin{table}

 \caption{\label{table:asymmetricruns}Summary of the parameters for the asymmetric runs}

 \begin{tabular}{|c|c|c|c|c|c|}
 \hline 
 \textbf{Name} & $\eta$ & $ \nu$  & $\Delta y$  & $x_m$ \\
 \hline
$A_{nd}^a$ & $2\ 10^{-3}$ &   $0$ &  $ 0.15$ & 300.\\
 \hline
$A_{nd}^b$ & $2\ 10^{-3}$ &   $0$ &  $ 0.076$ & $150$ \\
 \hline
$A_{\eta}^a$ & $1.4\ 10^{-2}$ &   $0$ &  $ 0.15$ & $150$\\
 \hline
$A_{\eta}^b$ & $1.4\ 10^{-2}$ &   $0$ &  $ 0.076$ & $150$\\
 \hline
$A_{\nu}^a$ & $2\ 10^{-3}$ &   $5\ 10^{-4}$ &  $ 0.076$ & $150$\\
 \hline
$A_{\nu}^b$ & $2\ 10^{-3}$ &   $5\ 10^{-4}$ &  $ 0.05$ & $100$\\
\hline
 \end{tabular}

 \end{table}

\subsection{Reconnection with numerical dissipation}
In this section, we analyze the results of runs $A_{nd}^{a,b}$. In these runs, the resistivity is so small that the reconnection of field line is enabled by numerical dissipation. The two runs differ by the mesh resolution. The resolution, in the $y$ direction, used for the runs $A_{nd}^a$  and $A_{nd}^b$ are $\Delta y=0.15$ and $\Delta y =0.076$, respectively. In the $x$ direction, both runs have the same resolution $\Delta x = 0.15$. These values are a factor of $\sqrt{3}$ smaller in terms of the ion inertial length based on the tenuous side of the current sheet. Consequently, the mesh is sufficiently fine to resolve the scales at which the Hall effect becomes important. To save computer time, the downstream length of the domain used for run $A_{nd}^b$ is set to $150$ while it is $300$ for run $A_{nd}^a$.  The Alfven time is about $\tau_A^a \approx 115$ for run $A_{nd}^a$ and $\tau_A^b\approx 60$ for run $A_{nd}^b$. The time step is $\Delta t=10^{-3}$ and $\Delta t=5\ 10^{-4}$ for $A_{nd}^a$ and $A_{nd}^b$, respectively. Such small time steps are necessary to prevent the violation of the CFL condition by the propagation of short wavelength whistler waves in the system. The width of the domain in the $y$ direction is set to $y_m=50$ for both runs. There are approximately $264$ millions macroparticles for each simulations.

\begin{figure}[htbp]
\begin{center}
\includegraphics[width=\linewidth]{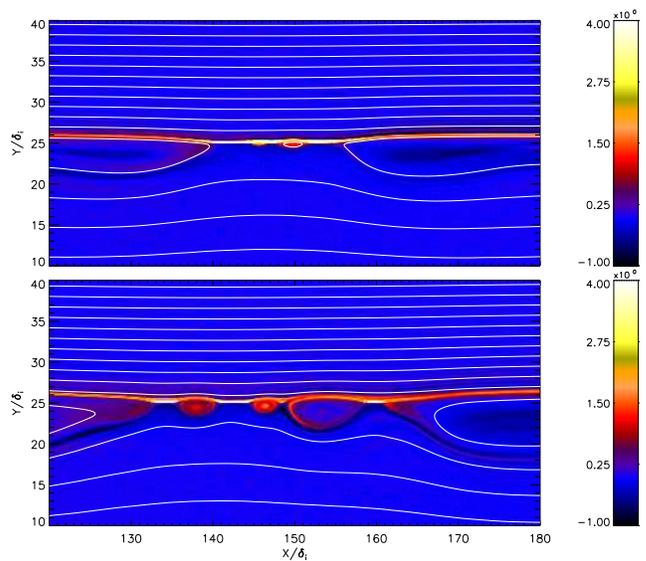}
\caption{Out-of-plane electron bulk flow $v_{ez}$ for the run $A_{nd}^a$ at $t=50$ (top) and $t=90$ (bottom). The in-plane magnetic field lines are represented by the solid white lines.}
\label{fig:cplotasym011}
\end{center}
\end{figure}

 Figure \ref{fig:cplotasym011} shows the out-of-plane electron bulk flow in simulation $A_{nd}^a$ at times $t=50$ (top panel) and $t=90$ (bottom panel). Surprisingly, the system starts with a very long period of time over which the current sheet elongates, while the flux tubes already reconnected by the initial perturbation are moving downstream. Then the reconnection process starts a globally unsteady phase and produces numerous plasmoids. After being born in the electron current sheet, these plasmoids grow quickly to ion scales and are slowly convected downstream. They dominantly grow on the weak field side owing to the weaker magnetic tension force there. Some plasmoids are ejected faster than others and coalesce with a preceding island.

\begin{figure}[htbp]
\begin{center}
\includegraphics[width=\linewidth]{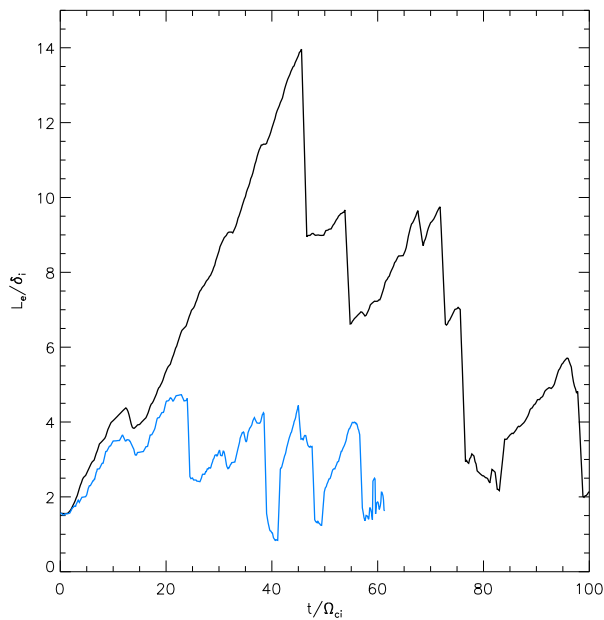}
\caption{Length of the electron current sheet as a function of time for the run $A_{nd}^a$ (black curve) and run $A_{nd}^b$ (blue curve).}
\label{fig:ecslengths}
\end{center}
\end{figure}

  As mentioned previously, plasmoids are seen in fully kinetic simulations of symmetric systems and are interpreted as the consequence of secondary tearing in long and thin electron current layers. Although it has different physical origins, this behavior is also reported in high-Lundquist number MHD simulations\citep{2009PhPl...16k2102B}. To understand further the origin of the plasmoids in our simulation, we therefore focus on the length of the electron current layer. To actually measure this length, we take the sum of the distances between the dominant X point and the points where the separatrices are distant from each other by more than a given threshold, here chosen equal to $0.7\delta_i$. Note that this measurement is not intended to be a quantitative measurement of the length of the current layer in itself, since it depends both on the arbitrary threshold and the angular aperture of the separatrices. It is however a simple measurement and a good qualitative proxy of the elongation of the current sheet and the resulting length visually corresponds to what would have been identified as the edges of the layer. The black curve on Fig. \ref{fig:ecslengths} is the result of this measure. As one can clearly see, the length of the layer increases rapidly to reach very high values and suddenly decreases at $t\approx 50$ after the formation of multiple plasmoids (also seen on the top panel of Fig. \ref{fig:cplotasym011}). It is interesting to notice that, all along the simulation, the current sheet does not stay short unless strongly influenced by two surrounding plasmoids. Their fast growth rate enables them to keep the current sheet short for a while before they move downstream. As they move away, the current sheet starts to elongate again with roughly the same speed until in breaks once again. Although the data shown here stops at $t=100$ because of a possible influence of the periodicity after the first Alfven time, the investigation of later times revealed similar behavior. The overall dynamics of magnetic reconnection in that system is thus dominated by an oscillation between an elongation of the electron current sheet and its breaking in multiple islands. 

The blue curve on Fig. \ref{fig:ecslengths} represents the length of the electron current layer for the run $A_{nd}^b$, where the mesh size has been divided by a factor $2$. We observe the same behavior although this time the current sheet elongates much less than in the previous case. Assuming the plasmoids are the result of a secondary (and numerical) tearing instability, the observation of shorter current sheets can be understood as a consequence of the use of a higher resolution mesh. Because the dissipation is explicitly neglected in the equations, the current sheet collapses to the first and only non-ideal scale, the grid scale. In the high resolution run, the current sheet will therefore be thinner than in the coarser case. Accordingly, the current sheet does not need to be as  long as in the coarse case to reach an aspect ratio where a numerical tearing is triggered and starts producing plasmoids. As a consequence of these shorter layers, instabilities occur more frequently but produce less plasmoids each time. One can conjecture that a run with a much higher resolution could then trigger a numerical tearing instability for very short electron layers. Depending on whether these islands manage to reach large scales or are quickly reconnected to exhaust field lines as they move downstream, the overall process could appear highly unsteady or spuriously steady macroscopically, with a very short electron layer.

\begin{figure}[htbp]
\begin{center}
\includegraphics[width=\linewidth]{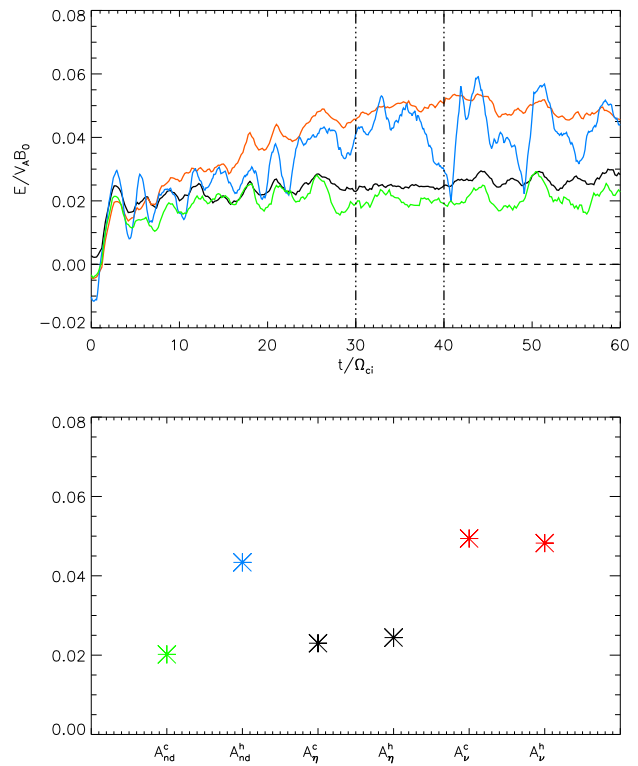}
\caption{\textbf{Top} : Time evolution of the reconnection rates for the runs $A_\nu^a$ (orange), $A_{nd}^b$ (blue), $A_{nd}^a$ (black) and $A_{\eta}^b$ (green). \textbf{Bottom} : Reconnection rate for all asymmetric runs, averaged between $t=30$ and $t=40$, as indicated by the dash-dotted vertical lines on the top panel.}
\label{fig:ratesasym}
\end{center}
\end{figure}

The top panel of Fig. \ref{fig:ratesasym} shows the time evolution of the reconnection rate for the runs $A_{nd}^a$ (green) and $A_{nc}^b$ (blue). The two curves are significantly different. The rate of the higher resolution run is much larger than the one of the coarse run, but also much less steady. In particular, strong oscillations can be observed around $t\approx 40$, $t\approx 50$, $t\approx 55$. These times corresponds to times where strong variations of the current sheet length are also observed. This suggests that the frequent formation of plasmoids, by maintaining the current layer short, plays an important role in controlling the reconnection rate. Whenever the formed plasmoids move far enough from their original location, the current sheet is less constrained and starts to elongate, and a decrease the reconnection rate is simultaneously observed (roughly, intervals $t\in [30,40]$, $t\in [43,48]$, $t \in [50,55]$). The very fast triggering of a plasmoid instability restores a faster rate, as opposed to the coarser resolution run. In the coarser case, the current layer does not break up into sufficiently small current sheets to maintain a fast rate. It is interesting to notice that this overall behavior is phenomenologically quite similar to interpretation recently proposed in symmetric fully kinetic models\citep{2007GeoRL..3413104K}, although here the instability must have a numerical origin.

We have performed two other simulations with the same resolution as run $A_{nd}^a$ but with smaller domain sizes in the downstream direction. In all cases, we clearly saw an elongation of the electron current sheet but the following evolution was interestingly quite different. In the smaller case, where $x_m=64$, we saw no plasmoid, whereas in the case where $x_m=150$, we saw the current sheet to break only once in multiple plasmoids and, after $t\approx 60$, stay short and localized at a single X point. Because the downstream size of the domain is the only difference between these runs, these features are the consequence of the periodicity influence on the system. In the smallest domain, the recirculation of plasma jets and electromagnetic fields occurs very soon. By artificially thickening the current layer, it prevents it from being unstable to a secondary tearing and plasmoids thus never form. In the intermediate size system, the same effect occurs although the system size being larger, it allowed for the first set of plasmoids to exist but does not let them enough time and space for the current sheet to have the opportunity to elongate again. The current sheet therefore stays short and the exhaust open, as a consequence of the periodic boundary and not because of a local mechanism. We observed the elongation/breaking mechanism to stop at $t\approx 60$, which approximately corresponds to one Alfven time for such system, consistently suggesting an influence of the domain periodicity. This influence might be less important for higher resolution runs since the current sheet is then always shorter.

\begin{figure}[htbp]
\begin{center}
\includegraphics[width=0.7\linewidth]{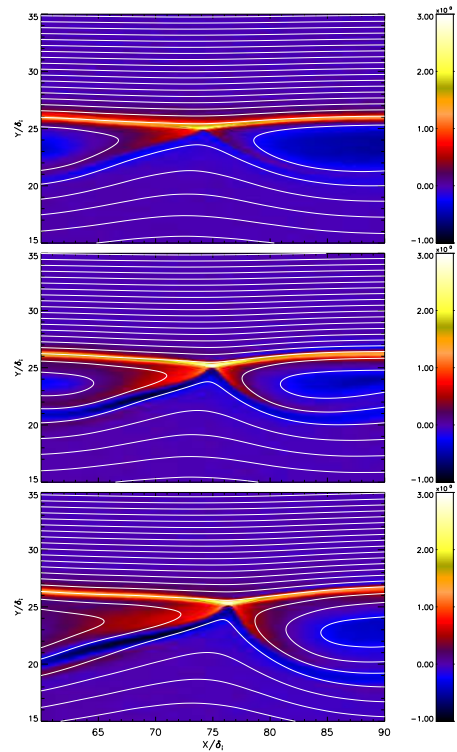}
\caption{On all panels, the color represents the out-of-plane electron bulk velocity $v_{ez}$ and the white solid lines are the in-plane magnetic field lines. \textbf{Top} :  run $A_\eta^{h}$ at time $t=50$. \textbf{Middle} : Run $A_{\nu}^{h}$ at time $t=34$ when the amount of reconnected flux is identical to the top panel. \textbf{Bottom} : Run $A_{\nu}^{h}$ at time $t=50$.}
\label{fig:cplotasym}
\end{center}
\end{figure}

\subsection{Reconnection with resistive dissipation}
As for the symmetric simulations, we now explore the consequences of increasing the value of the Joule resistivity up to $\eta=1.4\ 10^{-2}$. The size of the domain is set to $150$ in the downstream direction, and $50$ in the upstream direction. The resolution is set to $\Delta x = 0.15$ in both runs, while $\Delta y=0.15$ and $\Delta y=0.076$ for runs $A_\eta^a$ and $A_\eta^b$, respectively. The time step is set to $\Delta t = 10^{-3}$ for both simulations. There are approximately $132$ millions particles in each case.

\begin{figure}[htbp]
\begin{center}
\includegraphics[width=\linewidth]{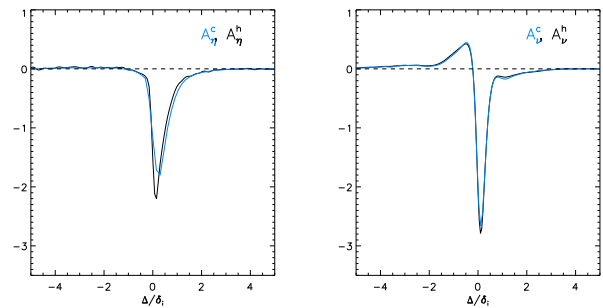}
\caption{$J_z$ measured along the $y$ direction through the X point and averaged between $t=30$ and $t=40$ to reduce numerical noise. \textbf{Left} : Runs $A_\eta^c$ (blue) and $A_\eta^h$ (black). \textbf{Right} : Runs $A_\nu^c$ (blue) and $A_\nu^h$ (right).}
\label{fig:jz}
\end{center}
\end{figure}

In the case of symmetric reconnection, increasing the resistivity to this value had no big overall effect on reconnection. The observed decrease of the reconnection rate was associated with the enhancement of large scale diffusion of electromagnetic energy and a small lengthening of the electron current layer. Surprisingly, increasing the resistivity in the asymmetric case now have important consequences regarding the large scale evolution of the system. As can be seen on the top panel of Fig. \ref{fig:cplotasym}, made from run $A_\eta^b$, the electron current sheet is short and localized. A visual inspection of the layer at different times throughout the simulation (not shown) revealed that this snapshot is a good representation of what occurs in the system at other (later) times. The run $A_\eta^{a}$ gave almost exactly the same result and is thus not shown here. As can be seen on the left panel of Fig. \ref{fig:jz}, the current sheet keeps approximately the same thickness when the resolution is doubled, which indicates that the Joule diffusion dominates the overall dissipation at the X line. However, the higher resolution case has a slightly higher current density, indicating again the difficulty of Joule diffusion to dissipate the incoming magnetic flux. The black curve on the top panel of Fig. \ref{fig:ratesasym} represents the reconnection rate for the run $A_\eta^b$. Consistently with the steady aspect of the current sheet, the reconnection rate is found to be rather constant, with a value around $\approx 0.025$. On the bottom panel of the same figure, one can see that both $A_\eta^a$ and $A_\eta^b$ have roughly the same reconnection rate.

\subsection{Reconnection with hyper-resistive dissipation}
Our last test consists in replacing the resistive term by a hyper-resistive term. In the symmetric case, hyper-resistivity was found to slightly decrease the rate compared to numerical dissipation, but the large scale features were observed to be unchanged, as opposed to the use of uniform Joule dissipation. In this asymmetric case, changing the dissipation physics again results in a different macroscopic behavior. The right panel of fig. \ref{fig:ecslengths} shows that for this value of hyper-resistivity $\nu=5\ 10^{-4}$, changing the resolution by a factor of $2$ leaves the current sheet unaffected, as opposed to the case of Joule diffusion where a small different can be noticed. The structure of the current layer is also significantly different and contrary to the resistive case, does not consist in a single peak anymore. Figure \ref{fig:cplotasym} shows the out-of-plane electron bulk velocity $v_{ez}$ for times $t=34.5$ (middle panel) and $t=50$ (bottom panel). The middle panel corresponds to a reconnected flux equivalent to the one of the top panel, in the resistive case $A_{\eta}^b$. At this same phase of the process, the two snapshots look globally similar, the electron current being locally enhanced at the reconnection site in both cases and the field lines dominantly bent on the weak field side. A more careful examination reveals differences. At the reconnection site as well as in the separatrix regions, the electron bulk flows are weaker in run $A_\eta$ than they are in run $A_\nu$. The left-right asymmetry resulting from the presence of an initial guide field seems to be more pronounced in the hyper-resistive run. Finally, the bottom panel of Fig. \ref{fig:cplotasym} shows that, at the same time, the reconnection process is much more advanced in the hyper-resistive case than it is in the resistive one. Consistently, the reconnection rate for the run $A_\nu^{a,b}$ is much larger than for $A_\eta^{a,b}$, as can be observed by comparing the black and orange curves on the top panel of fig. \ref{fig:cplotasym}. The bottom panel shows that the higher resolution run $A_\nu^b$ has the same reconnection rate.

\section{Summary and discussion}
We have performed two dimensional hybrid simulations to study the role of the dissipation scale physics in the process of magnetic reconnection and extend it to the more general case of asymmetric current layers. Standard hybrid models include ion kinetic physics and fluid electron physics but neglect electron inertial effects (thermal and bulk). These effects are indeed small at the ion scales, and are usually thought to have no dynamical role beyond violating the frozen-in condition. With the hybrid model, the use of different non-ideal electron mechanisms allowed us to investigate whether the nature of the dissipation occurring at the X line can significantly alter the large scale reconnection dynamics or not. In this paper, we have used numerical, resistive and hyper-resistive dissipation. \\

The first part of this study focused on a symmetric configuration, for which we have provided evidences that our model recovers basic features of 2D collisionless reconnection. The structure of the magnetic and electric Hall fields was found to be very similar to those measured in previous hybrid and fully kinetic calculations. Neglecting all dissipation mechanisms in Ohm's law, we observed the current layer to collapse down to the grid scale in both upstream and downstream directions, independently of the mesh size. However, as observed in previous studies, the overall reconnection rate seems unaffected and the process stays fast and steady, supporting the Hall reconnection paradigm. The use of a uniform resistivity increases the length of the current layer and also dissipates large scale electromagnetic energy in the exhaust region. As a result, the reconnection rate is significantly decreased. Peak amplitudes of electromagnetic fields at the ion scale are however unaffected by the viscous dissipation caused by hyper-resistivity, which remains at small scale, as previously understood. The reconnection rate is found to decrease a bit in comparison to the dissipationless case, however the difference remains small. A major difference between these results and modern fully kinetic calculations is the lack of plasmoids. In all cases here, the reconnection process was indeed steady. Whether this discrepancy means that hybrid models lack key physical ingredients is however still unclear, the formation mechanism of these plasmoids and its dependance on the inherent and over-estimated shot noise in Particle-In-Cell simulations being still unknown. Furthermore, if in situ measurements have provided evidences for the observation of such magnetic islands\citep{2007JGRA..11206235E,2008NatPh...4...19C}, the statistics of their occurrence his however unknown, leaving unanswered the question of their importance regarding the overall steadiness of magnetic reconnection in collisionless systems.  Fully kinetic Vlasov simulations can address this issue but remain computationally challenging considering the large domains required.\\

In a second part, we have implemented an asymmetric initial condition, already used in previous works and again performed several calculations with different non-ideal mechanisms. Contrary to the symmetric case, the nature of the non-ideal mechanism violating the frozen-in condition has been found to lead to substantial differences regarding the macroscopic reconnection dynamics. When the dissipation is neglected, the electron current layer is seen to elongate until it triggers a plasmoid instability. Because they grow faster than they move, the plasmoids constrain the current layer, which stays short for a while but starts elongating again as the islands move downstream. Consistently, a higher resolution simulation triggers the instability for shorter layers and leads to more frequent instabilities producing less plasmoids. As a result the current layer stays dynamically shorter and the reconnection rate is globally faster, but very unsteady because modulated by the island formation process. 

Surprisingly, the use of a uniform and quite large resistivity was observed to prevent the electron layer from elongating and no plasmoids were observed in such cases. Similarly, the hyper-resistivity keeps the electron layer short and the process steady. Like in the symmetric case, controlling the thickness of the current layer with a uniform resistivity without enabling large scale diffusion appears difficult.

This work suggests two major questions : 1/ Why is the electron layer elongating in the absence of physical dissipation in the Ohm's law while the Hall term is present? 2/ Why does it occur only in the asymmetric case?  Let us consider a case where fast reconnection occurs in a steady state, and at a given time the dissipation mechanism ceases for some reason. Accordingly, the field lines, in the upstream region, that were about to be reconnected cannot be reconnected anymore. In the exhaust however, the field lines that have already been reconnected will keep moving downstream and carry with them the hot plasma away from the reconnection region. As a result of these two effects, a pressure unbalance will force the localized electron layer to elongate and become a 1D tangential layer. If one turns the dissipation mechanism on again, new reconnected flux is provided to the exhaust, along which Hall disturbances can propagate and  open the exhaust again, increasing the reconnection rate. This interpretation is similar to the localizing effect of the non-ideal electron term, recently proposed\citep{2007GeoRL..3413104K,2009PhPl...16j2111S}. If numerical dissipation can violate the frozen-in condition and enable reconnection of field lines, one cannot, however, expect that it behaves like an actual dissipative effect, i.e. there is no reason for which it should have this localizing property and therefore an elongation of the current layer can be expected.

In the symmetric system however, the electron current sheet stays localized even in the absence of physical dissipation in the Ohm's law, which seems to contradict our previous interpretation. Let us notice the following facts: in this system, the exhaust is widely open on both upstream sides of the neutral line, as opposed to the asymmetric case where the field lines on the strong field side are barely bent. In the asymmetric dissipationless runs, we have noticed that two plasmoids growing rapidely around the X point can, for some time, maintain the electron layer short. Finally, we have also observed that the widening of the exhaust region due to the periodicity of the domain and the associated recirculation of plasma, can stop the elongation of the current layer and keep a single, localized, X point. Although further work is needed to fully understand this effect, these facts already suggest that the dissipation occurring at the X line may not be the only mechanism acting to constrain the length of the electron layer, and a feedback from the ion scale, whether it is controlled by the Hall effect, the growth of a nearby plasmoid or the widening effect associated to periodic recirculation, might also play an important role. The asymmetry of the initial system strongly affects the ion scale geometry surrounding the X line, which might then lead into a different constraint on the length of the electron current sheet than in symmetric systems.\\

In all the simulations presented in this work, the field line connectivity can be changed, whether this is due to grid effects, resistivity or hyper-resistivity. However, not all simulations show with the same overall evolution. Therefore, this work provides evidences that collisionless reconnection is generally not independent from the dissipation effects, as if they could be anything as long as they enable the change of connectivity. It is however possible that, for some reason that is yet not understood, the coupling of the Hall term with the appropriate dissipation term results in an evolution that, apparently, does not depend, or very weakly depends, on the parameter controlling the dissipation. Further work is needed to understand to what extent the reconnection rate depends on this localized dissipation once it is operating. This is beyond the scope of this paper and will be the topic of a future study. Preliminary results suggest that varying slightly the hyper-resistivity coefficient does not change the reconnection rate. Considering that, for these tests, smaller values of hyper-resistivity would be preferred to larger ones, the interval of values that can actually be tested is limited by the high resolution and small time steps required. These, with the requirement of long domains in the downstream direction, make such study difficult. Although hyper-resistiviy shares common properties with the kinetic dissipation in symmetric systems, it remains a simplified model and there is no theoretical proof that the similarity persists in asymmetric systems. Furthermore, the link between the non-uniform and self-consistent kinetic dissipation coefficient obtained from collisionless theory and the value of the uniform hyper-resistivity is not clear even in symmetric systems, therefore comparing the variation of the hyper-resistivity with the variation of the electron mass in a fully kinetic system is not trivial. Understanding the kinetic mechanisms leading to dissipation in asymmetric systems is thus an important issue and is crucial for the understanding of the future NASA Magnetospheric MultiScale (MMS) mission. Preliminary results in the comparison between hyper-resistive hybrid and fully kinetic PIC simulations of the same asymmetric systems reveal striking similarities and a detailed comparison is the topic of a separate paper\citep{aunai2013a}.













%




%











%







\begin{acknowledgments}

Three of us (N.A., C.B. and R.E.) acknowledge support from the NASA postdoctoral program. The authors acknowledge Roch Smets for the hybrid code.
\end{acknowledgments}


%

\end{document}